\documentclass[a4paper,twocolumn,11pt]{quantumarticle}
\pdfoutput=1
\usepackage[symbol]{footmisc}

\usepackage[utf8]{inputenc}
\usepackage[english]{babel}
\usepackage[T1]{fontenc}
\usepackage{amsmath}
\usepackage{hyperref}
\usepackage{breqn}
\usepackage{tikz}
\usepackage{lipsum}
\usepackage[numbers,sort&compress]{natbib}

\begin{document}

\title{Quantum Carleman Linearization of the Lattice Boltzmann Equation with Boundary Conditions}

\author{Bastien Bakker}
\affiliation{*HRL Labratories, Malibu, CA 90265, USA}
\author{Thomas W. Watts}
\maketitle
\begin{abstract}
 
 The Lattice Boltzmann Method (LBM) is widely recognized as an efficient algorithm for simulating fluid flows in both single-phase and multi-phase scenarios. In this research, a quantum Carleman Linearization formulation of the Lattice Boltzmann equation is described, employing the Bhatnagar Gross and Krook equilibrium function. Our approach addresses the treatment of boundary conditions with the commonly used bounce back scheme.

The accuracy of the proposed algorithm is demonstrated by simulating flow past a rectangular prism, achieving agreement with respect to fluid velocity in comparison to classical LBM simulations. This improved formulation showcases the potential to provide computational speed-ups in a wide range of fluid flow applications.

Additionally, we provide details on read in and read out techniques.
\end{abstract}

\section{Introduction}
Fluid flow simulations have myriad applications across engineering and scientific domains. One notably significant application pertains to fluid flow in porous media, which holds relevance for scenarios such as bone marrow circulation, hydrocarbon extraction from reservoirs, and filtration processes \cite{sharma2019lattice}. However, the simulation of porous media poses challenges, such as the characterization of pore-space geometry and the integration of fluid-solid interfacial phenomena \cite{sharma2019lattice}\cite{liu2016multiphase}. One possible technique to address these computational bottlenecks, the Lattice Boltzmann Method (LBM), has emerged as a promising contendor for addressing flow through porous media in view of its ability to handle complex boundary conditions and efficiently simulate fluid/fluid interfaces \cite{sharma2019lattice}. However, computational efficiency remains a concern, particularly when simulating large-scale porous structures, as these simulations have significant memory demands \cite{sharma2019lattice}, thereby motivating a quantum formulation of the LBM.

Several approaches have been established for a quantum Lattice Boltzmann (QLB) formulation, including the treatment of QLB as a quantum walk \cite{succi2015quantum} and the implementation of the LBM with quantum Carleman Linearization \cite{itani2022analysis}\cite{li2023potential}. While the Carleman Linearization technique applied to the LBM has showcased the potential for exponential quantum advantage in fluid dynamics simulations, the treatment of boundary conditions remains unaddressed \cite{li2023potential}. Thus, an improvement of the Quantum Carleman Linearization of the Lattice Boltzmann model (QCL-LBM) is necessary to treat boundary conditions.

This study introduces a QCL-LBM approach aimed at enhancing the capabilities of fluid flow simulations in quantum computing. The collision operator of the LBM is formulated similarly to previous methodologies employing the Bhatnagar Gross and Krook (BGK) equilibrium function, however, adaptations are made to the propagation step \cite{bhatnagar1954model}. This adapted step yields enhanced simulation accuracy and allows for treatment of boundary conditions using the widely adopted bounce back scheme \cite{succi2001lattice}.
\footnotetext{This material is based upon work supported by the Defense Advanced Research Projects Agency under Contract No. HR001122C0074. The support and program funded the entirety of this material.}

The objective of this paper is to demonstrate the precision and potential applications of the improved QCL-LBM, as well as describe read-in and read-out methods for a complete quantum circuit. To achieve this objective, the proposed formulation is benchmarked against a classical LBM simulation. As a means of validation, the flow around a rectangular prism is considered, a classical benchmark scenario in the realm of fluid flow simulations \cite{versteeg2007introduction}.

\section{Lattice Boltzmann Method}
The LBM is a powerful numerical technique used to simulate fluid flows. Originating from the ki-
netic theory of gases, the LBM has gained popularity due to its simplicity, efficiency, and ability to handle complex geometries \cite{chen1998lattice}.

At the core of the LBM lies the Lattice Boltzmann Equation (LBE), which provides a discrete, mesoscopic representation of the fluid dynamics governed by the Boltzmann equation \cite{he1997lattice}. The LBE describes the evolution of particle distribution functions, denoted as $f_{i}(\mathbf{x}, t)$, encoding the probability of finding particles with velocities $\mathbf{c}_{i}$ at a lattice node $\mathbf{x}$ and time $t$. A lattice is constructed from a spatial discretization, commonly done uniformly.

The evolution of these distribution functions over the lattice occurs through two fundamental steps: the collision step, where particles interact and redistribute their populations, and the streaming step, where particles propagate to neighboring lattice nodes based on their discrete velocities \cite{succi2001lattice}.

\subsection{Collision and Propagation}
The collision step is given by:
\begin{dmath}
f_{i}\left(\mathbf{x}, t+\delta_{t}\right)=f_{i}(\mathbf{x}, t)-\Omega\left(f_{i}(\mathbf{x}, t)\right)
\end{dmath}
where $\delta_{t}$ is the time step size and $\Omega\left(f_{i}(\mathbf{x}, t)\right)$ is the collision operator representing the interactions between particles. A common implementation of the collision operator is the BGK model which describes the fluid as relaxing towards equilibrium with collisions between the fluid molecules.

The equation for the BGK model is:
\begin{dmath}
\Omega\left(f_{i}(\mathbf{x}, t)\right)=\frac{\delta_{t}}{\tau}\left(f_{i}(\mathbf{x}, t)-f_{i}^{e q}(\mathbf{x}, t)\right)
\end{dmath}
where $\tau$ is the characteristic time scale and $f_{i}^{e q}(\mathbf{x}, t)$ is the equilibrium density calculated as follows:
\begin{dmath}
f_{i}^{e q}=w_{i} \rho\left[1+3 \mathbf{e}_{i} \cdot \mathbf{u}+\frac{9}{2}\left(\mathbf{e}_{i} \cdot \mathbf{u}\right)^{2}-\frac{3}{2} \mathbf{u} \cdot \mathbf{u}\right]
\end{dmath}
where $w_{i}$ is a weighting coefficient dependent upon the velocity scheme, $\mathbf{e}_{i}$ is the velocity direction vector associated with particle $i, \rho(\mathbf{x}, t)=$ $\sum_{i=0}^{Q} f_{i}(\mathbf{x}, t)$ is the fluid density at lattice point $\mathbf{x}$, $\mathbf{u}(\mathbf{x}, t)=\frac{1}{\rho(\mathbf{x}, t)} \sum_{i=0}^{Q} f_{i}(\mathbf{x}, t) \mathbf{e}_{i}$ is the macroscopic fluid velocity at point $\mathbf{x}$, and $Q$ is the number of discrete velocity directions, discussed below.

The propagation step is given by:
\begin{dmath}
f_{i}\left(\mathbf{x}+\mathbf{e}_{i}, t+\delta t\right)=f_{i}(\mathbf{x}, t)
\end{dmath}
In this step, the particle distribution functions are propagated towards nearby lattice points based upon their velocity.

\subsection{Common Velocity Schemes}
In the LBM, velocity is discretized into a fixed number of directions. The dimension the simulation is performed in and the number of velocity directions governs the velocity scheme, represented as $\mathrm{D} i \mathrm{Q} j$. Two widely used velocity schemes are D2Q9 and D3Q19, operating in two and three dimensions, respectively \cite{qian1992lattice}. D2Q9, shown in Figure 1, employs nine discrete velocities in a two-dimensional simulation, and is used in this paper.
\begin{figure}[t]
  \centering
  \includegraphics[width=150pt]{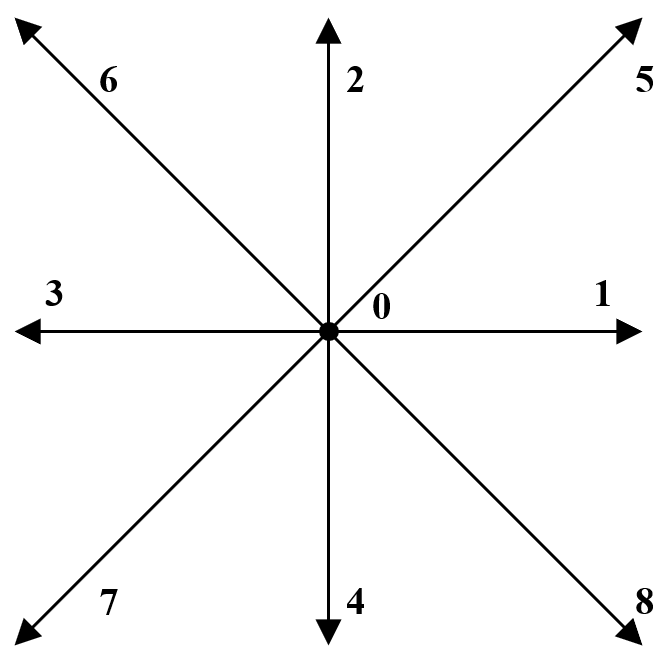}
  \caption{D2Q9 velocity scheme for LBM with numbers corresponding to index}
\end{figure}

\section{Carleman Linearization of the LBE}
In Carleman Linearization, a nonlinear dynamical system is transformed into an infinite dimensional linear system \cite{amini2021error}. The linear system solution $\phi^{k}$ is constructed from Kronecker products of the solution vector $f$ as such:
\begin{dmath}
\phi^{k}=\left(f, f^{\otimes 2}, \ldots, f^{\otimes k}\right)^{T}
\end{dmath}
and placed into a linear system:
\begin{dmath}
\frac{\partial \phi^{k}}{\partial t}=C^{k} \phi^{k}
\end{dmath}
where $C^{k}$ is derived from the nonlinear system associated with solution $f$. As $k$ approaches infinity, the solution to this system gives the exact solution of the nonlinear system:
\begin{dmath}
\frac{\partial \phi^{\infty}}{\partial t}=C^{\infty} \phi^{\infty}
\end{dmath}

\subsection{Scheme for D1Q3 model}
For simplicity, we first consider the D1Q3 scheme for LBM depicted in Figure 2. For the D1Q3 scheme, the weighting coefficients are:
\begin{figure}[t]
  \centering
  \includegraphics[width=180pt]{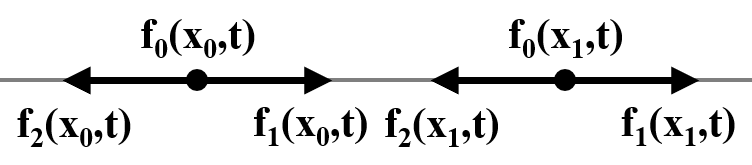}
  \caption{D1Q3 Grid with Two Lattice Points}
\end{figure}
\begin{dmath}
w=\left[\frac{2}{3}, \frac{1}{6}, \frac{1}{6}\right]
\end{dmath}
From equation 1, we derive the time evolution of $f_{0}\left(\mathbf{x}_{0}, t\right)$ for the collision step as:
\begin{dmath}
\frac{\partial f_{0}\left(\mathbf{x}_{0}, t\right)}{\partial t}  =\frac{1}{\tau}\left[f_{0}\left(\mathbf{x}_{0}, t\right)-\\
\frac{2}{3}\left(\rho-\frac{3}{2 \rho}\left(f_{1}\left(\mathbf{x}_{0}, t\right)-f_{2}\left(\mathbf{x}_{0}, t\right)\right)^{2}\right)\right]
\end{dmath}
where
\begin{dmath}
\rho=f_{0}\left(\mathbf{x}_{0}, t\right)+f_{1}\left(\mathbf{x}_{0}, t\right)+f_{2}\left(\mathbf{x}_{0}, t\right)
\end{dmath}
Since the distribution at index $i=0$ doesn't change in propagation, this is the final equation for the time step of $f_{0}\left(\mathbf{x}_{0}, t\right)$. To treat the $\frac{1}{\rho}$ prefactor, a first order Taylor approximation will be made as such:
\begin{dmath}
\frac{1}{\rho} \approx-\frac{1}{\bar{\rho}^{2}}(\rho-\bar{\rho})+\frac{1}{\bar{\rho}}
\end{dmath}
where $\bar{\rho}$ is the average fluid density. This approximation results in $O\left(M a^{4}\right)$ error, which is acceptable since the LBE approximates the Navier-Stokes Equations (NSE) to an error of $O\left(M a^{2}\right)$ \cite{li2023potential}.

For $f_{2}\left(\mathbf{x}_{0}, t\right)$, however, propagation must be taken into account. In a single time step, the particle population will be replaced by a collision adjusted population coming from lattice point $\mathbf{x}_{1}$. Specifically, $f_{2}\left(\mathbf{x}_{1}, t\right)$ will be propagated into its place. This can be represented as:
\begin{dmath}
f_{2}\left(\mathbf{x}_{0}, t+\delta_{t}\right)=f_{2}\left(\mathbf{x}_{1}, t\right)-f_{2}\left(\mathbf{x}_{0}, t\right)+\delta_{t} \Omega\left(f_{2}\left(\mathbf{x}_{1}, t\right)\right)
\end{dmath}
This is formulated differently than the previous QCL-LBM, with the populations being subtracted fully and replaced with propagated neighboring populations and the collision operator being dependent on that propagated population rather than the current population. 

\subsection{Carleman Linearization}
Since third order terms appear in the derivative of the first order terms, a truncation of at least order 3 is required to recover the NSE using Carleman Linearization \cite{li2023potential}. This can be represented as:
\begin{dmath}
\frac{\partial f}{\partial t}=S f+F_{1} f+F_{2} f^{\otimes 2}+F_{3} f^{\otimes 3}
\end{dmath}
where
\begin{dmath}
f(t)=\left(f_{0}\left(\mathbf{x}_{0}, t\right), f_{1}\left(\mathbf{x}_{0}, t\right), \ldots,\\ f_{0}\left(\mathbf{x}_{n}, t\right), \ldots, f_{Q}\left(\mathbf{x}_{n}, t\right)\right)^{T}
\end{dmath}
and $S$ denoting the streaming operator defined as:
\begin{dmath}
S f\left(\mathbf{x}_{i}, t\right)=\frac{1}{\delta_{t}}\left[f\left(\mathbf{x}_{i}-\mathbf{e}_{i}, t\right)-f\left(\mathbf{x}_{i}, t\right)\right]
\end{dmath}
Now, we may represent the collision and streaming operator in a Carleman Linearization matrix as such:
\begin{dmath}
C^{3}=\left(\begin{array}{ccc}
F_{1} & F_{2} & F_{3} \\
0 & F_{1}^{2} & F_{2}^{2} \\
0 & 0 & F_{1}^{3}
\end{array}\right)
\end{dmath}
and
\begin{dmath}
S^{3}=\left(\begin{array}{ccc}
S_{1} & 0 & 0 \\
0 & S_{1}^{2} & 0 \\
0 & 0 & S_{1}^{3}
\end{array}\right)
\end{dmath}
where
\begin{dmath}
F_{i}^{n}= F_{i} \otimes I^{\otimes(n-1)}+I \otimes F_{i} \otimes I^{\otimes(n-2)}+\ldots+I^{\otimes(n-1)} \otimes F_{i}
\end{dmath}
A larger truncation number may be used, but the resulting variable blowup would limit the number of nodes that are computationally tractable. The total number of variables associated with a truncation number $N$ can be represented as:
\begin{dmath}
\sum_{n=1}^{N}\left[\left(N_{f}-N_{B C}\right) \cdot Q\right]^{n}
\end{dmath}
where $N_{f}$ is the number of fluid nodes and $N_{B C}$ is the number of boundary condition nodes within simulation bounds. A forwards Euler approximation can now be made with $C^3$ and $S^3$ inserted into the linear system equation:
\begin{dmath}
A=\left(\begin{array}{cccc}I & 0 & \cdots & 0\\ -\hat{O} & I & \cdots & 0 \\ \vdots & \ddots & \ddots & \vdots \\ 0 & 0 & -\hat{O}& I\end{array}\right)
\end{dmath}
where 
\begin{dmath}
\hat{O} = \left[I+\delta_{t}(C^3+S^3)\right]
\end{dmath}
and
\begin{dmath}
A\left(\begin{array}{c}
\phi\left(t_{0}\right) \\
\phi\left(t_{1}\right) \\
\vdots \\
\phi\left(t_{n}\right)
\end{array}\right)=\left(\begin{array}{c}
\phi\left(t_{0}\right) \\
0 \\
\vdots \\
0
\end{array}\right)
\end{dmath}
where $\phi(t_n)$ is the solution $\phi$ at the $n$-th time step, $\phi(t_0)$ is the initial condition generated as:
\begin{dmath}
\phi(t_0) = \begin{bmatrix}f(t_0)\\f(t_0)^{\otimes 2}\\f(t_0)^{\otimes 3}\end{bmatrix}
\end{dmath}

\subsection{Boundary Conditions}
\begin{figure}[t]
  \centering
  \includegraphics[width=190pt]{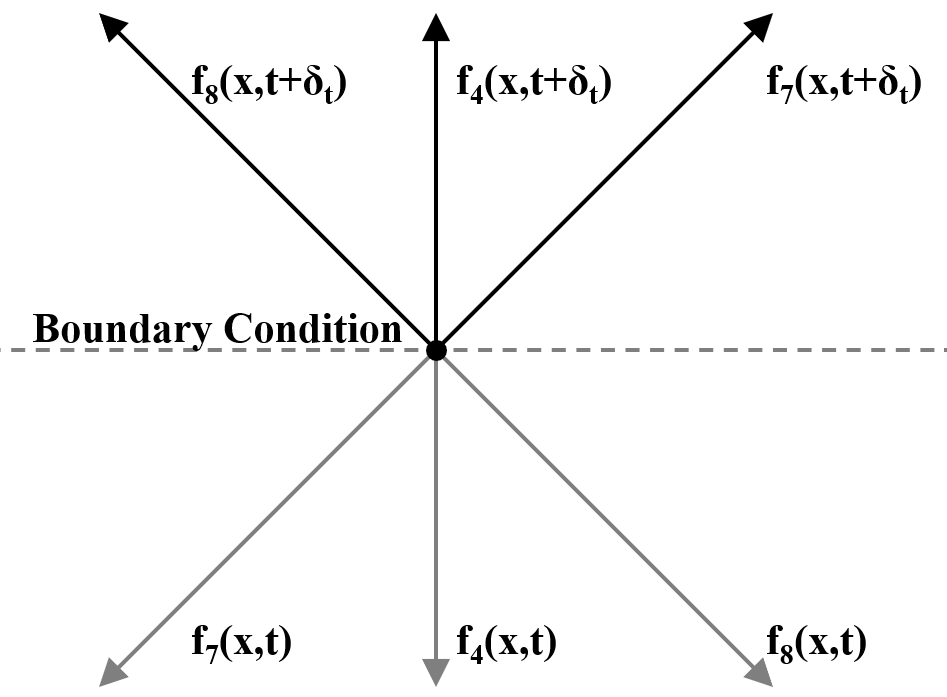}
  \caption{Bounce back scheme for LBM}
\end{figure}
Treatment of the interaction of fluid with solid boundary conditions is important to the accuracy of the LBM. The bounce back scheme is commonly used to impose no-slip boundary conditions, particularly for solid walls \cite{ladd1993short}. In this method, when particles collide with a boundary, their velocities are reversed. This results in a modified propagation step, depicted in Figure 3.

Using the bounce back scheme, boundary conditions can be implemented under this streaming operator construction. To demonstrate, the example from Figure 4 will be worked out.
\begin{figure}[t]
  \centering
  \includegraphics[width=180pt]{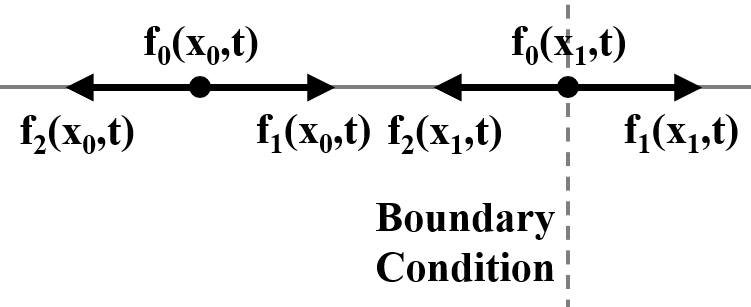}
  \caption{D1Q3 Grid with Two Lattice Points and
Boundary Condition}
\end{figure}
In a time step, populations that propagate into boundaries "bounce" back and adopt an opposing velocity. Imposing the bounce back scheme will result in the distribution $f_{1}\left(\mathbf{x}_{1}, t\right)$ reflecting back towards the population of $f_{2}\left(\mathbf{x}_{1}, t\right)$. As such the time stepping equation can be written out as:
\begin{dmath}
f_{2}\left(\mathbf{x}_{1}, t+\delta_{t}\right)=f_{1}\left(\mathbf{x}_{1}, t\right)-f_{2}\left(\mathbf{x}_{1}, t\right)+\delta_{t} \Omega\left(f_{1}\left(\mathbf{x}_{1}, t\right)\right)
\end{dmath}
To implement this, the $C$ and $S$ operators must be modified. To do so, a pre-streaming step is performed, determining the resulting location of various populations after a classical streaming step. The results of this step are then used to shift terms in $C$ with various swaps making the operator dependent upon the correct populations. This entails using the streaming step of the imaginary node outside the boundaries and replacing it with the reflected population. To create $S$, a similar process is performed. Since the streaming operator defined in Equation 15 depends on itself regardless of streamed population, the identity matrix is adjusted with identical swaps to represent the $f(\mathbf{x}_i-\mathbf{e}_i,t)$ term, and then added to a negative identity matrix, representing the $-f(\mathbf{x}_i,t)$ term. This allows the population to be adjusted by subtracting the current population and adding the new streamed population. Modifying these two operators allows for the QCL-LBM to properly treat boundary conditions both within and on the bounds of the simulation.

\section{Computational Results}
To test the improved schema for the QCL-LBM, code was developed to process the collision and streaming matrices. These can then be placed into the large A matrix formulated in Equation 20. However, a complete quantum computer would be required to solve the linear system, thus to simulate, we iteratively step the solution through time. This was done with a D2Q9 formulation to test flow past a rectangular prism, although extending the method to 3 dimensions is trivial. The solution was then tested against a classical simulation of Lattice Boltzmann with respect to total fluid velocity to verify the effectiveness of the method.

This was done on a grid of size $10 \times 5$ nodes with a Reynolds number of 50 . An initial velocity $u$ was enforced by weighting the initial distributions of the nodes such that their fluid velocity acted towards the right:
\begin{dmath}
f_{\text {init }}=\bar{\rho}\left(\frac{4}{9}, \frac{1}{9}+\frac{u}{2}, \frac{1}{9}, \frac{1}{9}-\frac{u}{2}, \frac{1}{9}, \frac{1}{36}, \frac{1}{36}, \frac{1}{36}, \frac{1}{36}\right)
\end{dmath}
The initial condition was applied to all nodes to mimic common flow past objects simulations \cite{li2004numerical}.

No slip boundary conditions implemented using bounce back were placed on the y direction walls of the grid, while periodic boundary conditions were placed on the $\mathrm{x}$ direction walls to model flow towards the right. This was done over a timescale of 5 seconds and a step size of .00025, with results shown in Figure 5. A boundary condition was included at point $\mathrm{x}=2, \mathrm{y}=2$ to model model a square object and an initial velocity of 0.0001 lattice units per time step was enforced in the positive $\mathrm{x}$ direction.
\begin{figure}[t]
  \centering
  \includegraphics[width=250pt]{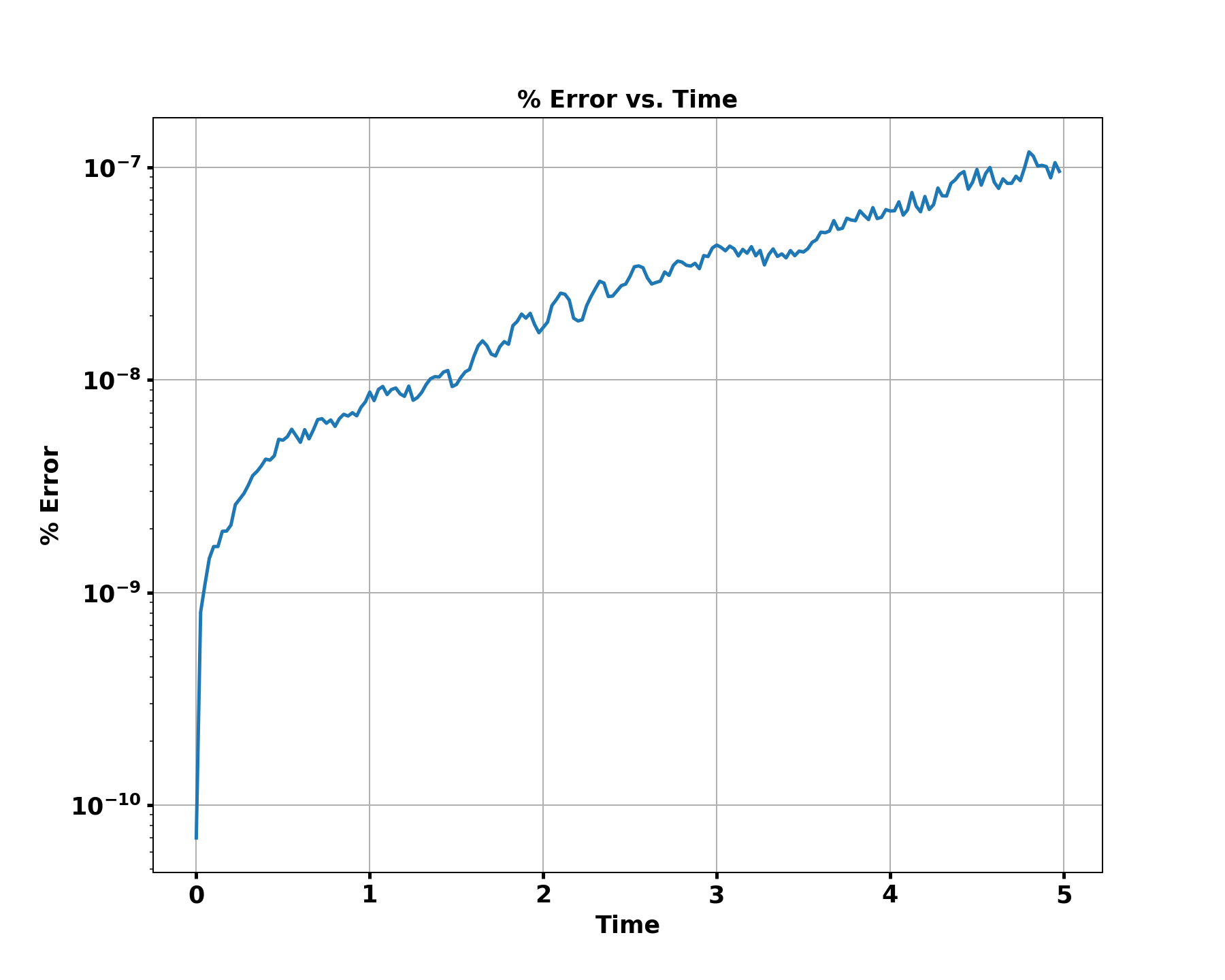}
  \caption{Percent error of total fluid velocity for quantum simulation compared to classical simulation}
\end{figure}
These simulations demonstrate the ability of the proposed QCL-LBM model to achieve agreement with a classical LBM simulation, despite sacrificing the exactness of streaming \cite{itani2022analysis}.

\subsection{Read Out}
In order to run a useful quantum computation, the problems of read in and read out must be addressed. An initial condition can be applied to all lattice nodes creating the input vector algorithmically, and the $A$ matrix can be constructed as described above. The linear system can then be solved for the solution vector with various quantum linear solvers \cite{childs2017quantum}\cite{costa2022optimal}. However, decoding quantum states into a classical vector normally removes quantum advantage, requiring some extractable metric to remain efficient \cite{zhang2021quantum}. Hence, the extraction of a sub-exponential number of numerical quantities from the large solution vector is integral to maintain speedup. We propose the extraction of the drag coefficient, a commonly extracted measure for porous media simulation\cite{wittig2017drag}\cite{patursson2010development}. The drag coefficient can be calculated by\cite{kajzer2017application}:
\begin{dmath}
C_{D}=2 \frac{F_{x}}{\rho U_{l}^{2} S}
\end{dmath}
where $U_{l}$ is the inflow speed in lattice units, $S$ is the cross section area of the boundary condition in lattice units, and $F_{x}$ is the force acting on the boundary condition calculated with\cite{kajzer2017application}:
\begin{dmath}
F=\sum_{\mathbf{x}_{s f}} \sum_{i}\left(f_{-i}\left(\mathbf{x}_{s f}, t\right)+f_{i}\left(\mathbf{x}_{s f}, t+\delta t\right)\right) \mathbf{e}_{i}
\end{dmath}
where $\mathbf{x}_{s f}$ are the fluid lattice units which have solid neighbors, and the subscript $-i$ is defined by $\mathbf{e}_{-i}=-\mathbf{e}_{i}$.

The drag coefficient can be extracted from the resultant solution vector of the linear system solver, giving a useful read out value of the quantum simulation.

\section{Conclusion}
The presented improvement to the QCL-LBM framework, with a modified formulation to increase accuracy and account for boundary conditions, results in high accuracy fluid simulation and enables integration of boundary conditions. This broadens the utility of the quantum Lattice Boltzmann approach and addresses memory constraints inherent in complex porous media scenarios. Additionally, we introduced a state preparation and read out technique, facilitating the construction of a complete quantum circuit. Further research is necessary to explore diverse approaches for handling boundary conditions, thereby extending the method's applicability even further.

\bibliographystyle{unsrtnat}
\bibliography{mybib}{}

\begin{thebibliography}{20}
\providecommand{\natexlab}[1]{#1}
\providecommand{\url}[1]{\texttt{#1}}
\expandafter\ifx\csname urlstyle\endcsname\relax
  \providecommand{\doi}[1]{doi: #1}\else
  \providecommand{\doi}{doi: \begingroup \urlstyle{rm}\Url}\fi

\bibitem[Sharma et~al.(2019)Sharma, Straka, and Tavares]{sharma2019lattice}
Keerti~Vardhan Sharma, Robert Straka, and Frederico~Wanderley Tavares.
\newblock Lattice boltzmann methods for industrial applications.
\newblock \emph{Industrial \& Engineering Chemistry Research}, 58\penalty0 (36):\penalty0 16205--16234, 2019.

\bibitem[Liu et~al.(2016)Liu, Kang, Leonardi, Schmieschek, Narv{\'a}ez, Jones, Williams, Valocchi, and Harting]{liu2016multiphase}
Haihu Liu, Qinjun Kang, Christopher~R Leonardi, Sebastian Schmieschek, Ariel Narv{\'a}ez, Bruce~D Jones, John~R Williams, Albert~J Valocchi, and Jens Harting.
\newblock Multiphase lattice boltzmann simulations for porous media applications: A review.
\newblock \emph{Computational Geosciences}, 20:\penalty0 777--805, 2016.

\bibitem[Succi et~al.(2015)Succi, Fillion-Gourdeau, and Palpacelli]{succi2015quantum}
Sauro Succi, Fran{\c{c}}ois Fillion-Gourdeau, and Silvia Palpacelli.
\newblock Quantum lattice boltzmann is a quantum walk.
\newblock \emph{EPJ Quantum Technology}, 2\penalty0 (1):\penalty0 1--17, 2015.

\bibitem[Itani and Succi(2022)]{itani2022analysis}
Wael Itani and Sauro Succi.
\newblock Analysis of carleman linearization of lattice boltzmann.
\newblock \emph{Fluids}, 7\penalty0 (1):\penalty0 24, 2022.

\bibitem[Li et~al.(2023)Li, Yin, Wiebe, Chun, Schenter, Cheung, and M{\"u}lmenst{\"a}dt]{li2023potential}
Xiangyu Li, Xiaolong Yin, Nathan Wiebe, Jaehun Chun, Gregory~K Schenter, Margaret~S Cheung, and Johannes M{\"u}lmenst{\"a}dt.
\newblock Potential quantum advantage for simulation of fluid dynamics.
\newblock \emph{arXiv preprint arXiv:2303.16550}, 2023.

\bibitem[Bhatnagar et~al.(1954)Bhatnagar, Gross, and Krook]{bhatnagar1954model}
Prabhu~Lal Bhatnagar, Eugene~P Gross, and Max Krook.
\newblock A model for collision processes in gases. i. small amplitude processes in charged and neutral one-component systems.
\newblock \emph{Physical review}, 94\penalty0 (3):\penalty0 511, 1954.

\bibitem[Succi(2001)]{succi2001lattice}
Sauro Succi.
\newblock \emph{The lattice Boltzmann equation: for fluid dynamics and beyond}.
\newblock Oxford university press, 2001.

\bibitem[Versteeg and Malalasekera(2007)]{versteeg2007introduction}
Henk~Kaarle Versteeg and Weeratunge Malalasekera.
\newblock \emph{An introduction to computational fluid dynamics: the finite volume method}.
\newblock Pearson education, 2007.

\bibitem[Chen and Doolen(1998)]{chen1998lattice}
Shiyi Chen and Gary~D Doolen.
\newblock Lattice boltzmann method for fluid flows.
\newblock \emph{Annual review of fluid mechanics}, 30\penalty0 (1):\penalty0 329--364, 1998.

\bibitem[He and Doolen(1997)]{he1997lattice}
Xiaoyi He and Gary Doolen.
\newblock Lattice boltzmann method on curvilinear coordinates system: flow around a circular cylinder.
\newblock \emph{Journal of Computational Physics}, 134\penalty0 (2):\penalty0 306--315, 1997.

\bibitem[Qian et~al.(1992)Qian, d'Humi{\`e}res, and Lallemand]{qian1992lattice}
Yue-Hong Qian, Dominique d'Humi{\`e}res, and Pierre Lallemand.
\newblock Lattice bgk models for navier-stokes equation.
\newblock \emph{Europhysics letters}, 17\penalty0 (6):\penalty0 479, 1992.

\bibitem[Amini et~al.(2021)Amini, Sun, and Motee]{amini2021error}
Arash Amini, Qiyu Sun, and Nader Motee.
\newblock Error bounds for carleman linearization of general nonlinear systems.
\newblock In \emph{2021 Proceedings of the Conference on Control and its Applications}, pages 1--8. SIAM, 2021.

\bibitem[Ladd(1993)]{ladd1993short}
Anthony~JC Ladd.
\newblock Short-time motion of colloidal particles: Numerical simulation via a fluctuating lattice-boltzmann equation.
\newblock \emph{Physical Review Letters}, 70\penalty0 (9):\penalty0 1339, 1993.

\bibitem[Li et~al.(2004)Li, Shock, Zhang, and Chen]{li2004numerical}
Yanbing Li, Richard Shock, Raoyang Zhang, and Hudong Chen.
\newblock Numerical study of flow past an impulsively started cylinder by the lattice-boltzmann method.
\newblock \emph{Journal of Fluid Mechanics}, 519:\penalty0 273--300, 2004.

\bibitem[Childs et~al.(2017)Childs, Kothari, and Somma]{childs2017quantum}
Andrew~M Childs, Robin Kothari, and Rolando~D Somma.
\newblock Quantum algorithm for systems of linear equations with exponentially improved dependence on precision.
\newblock \emph{SIAM Journal on Computing}, 46\penalty0 (6):\penalty0 1920--1950, 2017.

\bibitem[Costa et~al.(2022)Costa, An, Sanders, Su, Babbush, and Berry]{costa2022optimal}
Pedro~CS Costa, Dong An, Yuval~R Sanders, Yuan Su, Ryan Babbush, and Dominic~W Berry.
\newblock Optimal scaling quantum linear-systems solver via discrete adiabatic theorem.
\newblock \emph{PRX Quantum}, 3\penalty0 (4):\penalty0 040303, 2022.

\bibitem[Zhang et~al.(2021)Zhang, Hsieh, Liu, and Tao]{zhang2021quantum}
Kaining Zhang, Min-Hsiu Hsieh, Liu Liu, and Dacheng Tao.
\newblock Quantum gram-schmidt processes and their application to efficient state readout for quantum algorithms.
\newblock \emph{Physical Review Research}, 3\penalty0 (4):\penalty0 043095, 2021.

\bibitem[Wittig et~al.(2017)Wittig, Nikrityuk, and Richter]{wittig2017drag}
Kay Wittig, Petr Nikrityuk, and Andreas Richter.
\newblock Drag coefficient and nusselt number for porous particles under laminar flow conditions.
\newblock \emph{International Journal of Heat and Mass Transfer}, 112:\penalty0 1005--1016, 2017.

\bibitem[Patursson et~al.(2010)Patursson, Swift, Tsukrov, Simonsen, Baldwin, Fredriksson, and Celikkol]{patursson2010development}
{\O}ystein Patursson, M~Robinson Swift, Igor Tsukrov, Knud Simonsen, Kenneth Baldwin, David~W Fredriksson, and Barbaros Celikkol.
\newblock Development of a porous media model with application to flow through and around a net panel.
\newblock \emph{Ocean Engineering}, 37\penalty0 (2-3):\penalty0 314--324, 2010.

\bibitem[Kajzer and Pozorski(2017)]{kajzer2017application}
Adam Kajzer and Jacek Pozorski.
\newblock Application of the lattice boltzmann method to the flow past a sphere.
\newblock \emph{Journal of Theoretical and Applied Mechanics}, 55\penalty0 (3):\penalty0 1091--1099, 2017.

\end{thebibliography}
\end{document}